\documentclass[aoas,preprint]{imsart}

\RequirePackage[OT1]{fontenc}
\RequirePackage{amsthm,amsmath}
\RequirePackage[authoryear]{natbib}
\RequirePackage[colorlinks,citecolor=blue,urlcolor=blue]{hyperref}
\RequirePackage{bm}
\RequirePackage{amssymb}
\RequirePackage{booktabs}
\RequirePackage{adjustbox}
\RequirePackage{subfigure}
\RequirePackage{caption}
\RequirePackage{float}
\RequirePackage{listings}
\RequirePackage{amstext}
\RequirePackage{mathpazo}
\RequirePackage[scaled=0.92]{helvet}
\RequirePackage{courier}
\RequirePackage[T1]{fontenc}
\RequirePackage[latin9]{inputenc}
\RequirePackage[a4paper]{geometry}

\arxiv{arXiv:1702.05008}

\startlocaldefs
\numberwithin{equation}{section}
\theoremstyle{plain}

\endlocaldefs

\begin{document}

\begin{frontmatter}
\title{Tree ensembles with rule structured horseshoe regularization}
\runtitle{Tree ensembles with horseshoe regularization}

\begin{aug}
\author{\fnms{Malte} \snm{Nalenz}\ead[label=e1]{malte.nlz@googlemail.com}} \and
\author{\fnms{Mattias} \snm{Villani}\ead[label=e2]{mattias.villani@liu.se}\ead[label=u1,url]{http://mattiasvillani.com}}

\runauthor{M. Nalenz and M. Villani}

\affiliation{German Research Center for Environmental Health and Link\"oping University}

\address{Malte Nalenz\\
Helmholtz Zentrum M\"unchen\\
German Research Center for \\ Environmental Health \\
Institute of Computational Biology\\
Ingolst\"adter Landstra{\ss}e 1 \\
85764 Neuherberg, Germany \\
\printead{e1}\\
\phantom{E-mail:\ }}

\address{Mattias Villani \\ 
Division of Statistics and Machine Learning \\
Dept. of Computer and Information Science \\
SE-581 83 Link\"oping, Sweden \\
\printead{e2}\\
\printead{u1}}
\end{aug}

\begin{abstract}
Abstract. We propose a new Bayesian model for flexible nonlinear regression and classification using tree ensembles. The model is based on the RuleFit approach in \cite{RuleFit} where rules from decision trees and linear terms are used in a L1-regularized regression. We modify RuleFit by replacing the L1-regularization by a horseshoe prior, which is well known to give aggressive shrinkage of noise predictors while leaving the important signal essentially untouched. This is especially important when a large number of rules are used as predictors as many of them only contribute noise. Our horseshoe prior has an additional hierarchical layer that applies more shrinkage a priori to rules with a large number of splits, and to rules that are only satisfied by a few observations. The aggressive noise shrinkage of our prior also makes it possible to complement the rules from boosting in RuleFit with an additional set of trees from Random Forest, which brings a desirable diversity to the ensemble. We sample from the posterior distribution using a very efficient and easily implemented Gibbs sampler. The new model is shown to outperform state-of-the-art methods like RuleFit, BART and Random Forest on 16 datasets. The model and its interpretation is demonstrated on the well known Boston housing data, and on gene expression data for cancer classification. The posterior sampling, prediction and graphical tools for interpreting the model results are implemented in a publicly available R package. 
\end{abstract}



\end{frontmatter}

\section{Introduction}

Learning and prediction when the mapping between input and outputs
is potentially nonlinear and observed in noise remains a major challenge.
Given a set of $N$ training observations $(\mathbf{x},y)_{i},i=1,\dots,N$,
we are interested in learning or approximating an unknown function
$f$ observed in additive Gaussian noise
\[
y=f(\mathbf{x})+\epsilon,\;\;\epsilon\sim\mathcal{N}(0,\sigma^{2}),
\]
and to use the model for prediction. A popular approach is to use
a learning ensemble \citep{breiman1996stacked, Breiman2001, freund1996experiments, friedman2000greedy}
\[
f(\mathbf{x})=\sum_{l=1}^{m}\alpha_{l}f_{l}(\mathbf{x}),
\]
where $f_{l}(\mathbf{x})$ is a basis function (also called a weak
learner in the machine learning literature) for a subset of the predictors.
A variety of basis functions $f_{l}$ have been proposed in the last
decades, and we will here focus on decision rules. Decision rules
are defined by simple if-else statements and therefore highly interpretable
by humans. Finding a set of optimal rules is NP hard \citep{RuleFit},
and most practical algorithms therefore use a greedy learning procedure.
Among the most powerful are divide and conquer approaches \citep{ripper,furnkranz1999separate}
and boosting \citep{schapire99brief,dembczynski2010ender}.

A new way to learn decision rules is introduced in
\cite{RuleFit} in their RuleFit approach. RuleFit is
estimated by a two-step procedure. The \emph{rule generation} step
extracts decision rules from an ensemble of trees trained with gradient
boosting. The second \emph{regularizaton} step learns the weights
$\alpha_{l}$ for the generated rules via L1-regularized (Lasso) regression,
along with weights on linear terms included in the model. This is similar to stacking \citep{wolpert1992stacked, breiman1996stacked}, with the important difference that the members of the ensemble are not learned decision trees or other predictors, but individual rules extracted from trees. As argued in \cite{RuleFit}, this makes RuleFit a more interpretable model and, we argue below, has important consequences for the regularization part. RuleFit
has been successfully applied in particle physics, in medical informatics
and in life sciences. Our paper makes the following contributions
to improve and enhance RuleFit. 

First, we replace the L1-regularization \citep{Tibshirani1996} in
RuleFit by a generalized horseshoe regularization prior \citep{carvalho2010horseshoe}
tailored specifically to covariates from a rule generation step. L1-regularization
is computationally attractive, but has the well known drawback of
also shrinking the effect of the important covariates. This is especially
problematic here since the number of rules from the rule generation
step can be very large while potentially only a small subset is necessary
to explain the variation in the response. Another consequence of the
overshrinkage effect of the L1-regularization is that it is hard to
choose an optimal number of rules; increasing the number of rules
affects the shrinkage properties of the Lasso. This makes it very
hard to determine the number of rules a priori, and one has to resort
to cross-validation, thereby mitigating the computational advantage
of the Lasso. A horseshoe prior is especially attractive for rule
learning since it shrinks uninformative predictors aggressively while
leaving important ones essentially untouched. Inspired by the prior
distribution on the tree depth in Bayesian Additive Regression Trees (BART) \citep{0806.3286v2}, we design
a generalized horseshoe prior that shrinks overly complicated and
specific rules more heavily, thereby mitigating problems with overfitting.
This is diametrically opposed to RuleFit, and to BART and boosting,
which all combine a myriad of rules into a collective where single rules
only play a very small part.

Second, we complement the tree ensemble from gradient boosting \citep{friedman2000greedy}
in RuleFit with an additional set of trees generated
with Random Forest. The error-correcting nature of boosting makes
the rules highly dependent on each other. Trees from Random Forest
\citep{Breiman2001} are much more random and adding them to rules
from boosting therefore brings a beneficial diversity to the tree
ensemble. Note that it is usually not straightforward to combine individual trees from different ensemble strategies in a model; our combination of RuleFit and horseshoe regularization is an ideal setting for mixing ensembles since RuleFit makes it easy to combine ensembles, and the horseshoe prior can handle a large number of noise rules without overfitting.

Third, an advantage of our approach compared to many other flexible
regression and classification models is that predictions from our
model are based on a relatively small set of interpretable decision
rules. The possibility to include linear terms also simplifies interpretation
since it avoids a common problem with decision trees that linear relationships
need to be approximated with a large number of rules. To further aid
in the interpretation of the model and its predictions, we also propose
graphical tools for analyzing the model output. We also experiment
with post-processing methods for additional pruning of rules to simplify
the interpretation even further using the method in \cite{hahn2015decoupling}.

We call the resulting two-step procedure with mixed rule generation
followed by generalized rule structured horseshoe regularization the
\emph{HorseRule} model. We show that HorseRule's ability to keep the
important rules and aggressively removing unimportant noise rules
leads to both great predictive performance and high interpretability. 

The structure of the paper is as follows. Section 2 describes the
decision rule generation method in HorseRule. Section 3 presents
the horseshoe regularization prior and the MCMC algorithm for posterior
inference. Section 4 illustrates aspects of the approach on simulated
data and evaluates and compares the predictive performance of HorseRule
to several main competing methods on a wide variety of real datasets.
Section 5 concludes.

\section{Decision Rule Generation}

This section describes the \emph{rule generation} \emph{step} of HorseRule,
which complements the rules from gradient boosting in \citet{RuleFit}
with rules from Random Forest with completely different properties.

  \begin{minipage}{\textwidth}   \begin{minipage}[b]{0.49\textwidth}       \centering     \includegraphics[width=7cm, height=7cm]{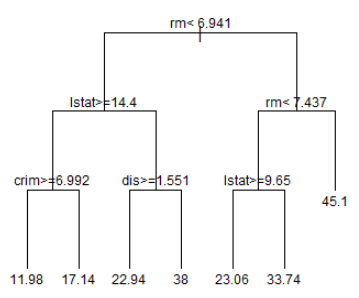}     \captionof{figure}{Decision Tree for the Boston Housing data.}     \label{tree}   \end{minipage}   \hfill   \begin{minipage}[b]{0.49\textwidth}     \centering     \begin{adjustbox}{center, width=6cm, height=3.15cm}     \begin{tabular}{cc}\hline       Rules & Conditions \\ \hline         $r_1$ & $RM \geq 6.94$ \\         $r_2$ & $RM < 6.94$ \\         $r_3$ & $RM < 6.94 \;\&\; LSTAT < 14.4$ \\         $r_4$ & $RM < 6.94 \;\&\; LSTAT \geq 14.4$ \\         $r_5$ & $RM < 6.94 \;\&\; LSTAT < 14.4 \;\&\; CRIM<6.9$ \\         $r_6$ & $RM < 6.94 \;\&\; LSTAT < 14.4 \;\&\; CRIM \geq 6.9$ \\         $r_7$ & $RM \geq 6.94 \;\&\; LSTAT < 14.4 \;\&\; DIS < 1.5$ \\         $r_8$ & $RM \geq 6.94 \;\&\; LSTAT < 14.4 \;\&\; DIS \geq 1.5$ \\         $r_9$ & $6.94 \leq RM < 7.45$ \\         $r_{10}$ & $6.94 \leq RM < 7.45$\\         $r_{11}$ & $6.94 \leq RM < 7.45 \;\&\; LSTAT < 9.7$ \\         $r_{12}$ & $6.94 \leq rm < 7.45 \;\&\; LSTAT \geq 9.7$ \\         \hline       \end{tabular}       \label{rules}       \end{adjustbox}       \captionof{table}{Corresponding rules, defining the Decision Tree.}     \end{minipage}   \end{minipage}

\subsection{Decision Rules}\label{sec:rule-generation}

Let $S_{k}$ denote the set of possible values of the covariate $x_{k}$
and let $s_{k,m}\subseteq S_{k}$ denote a specific subset. A decision
rule can then be written as 
\begin{equation}
r_{m}(\bm{x})=\prod_{k\in Q_{m}}I(x_{k}\in s_{k,m}),\label{rule}
\end{equation}
where $I(x)$ is the indicator function and $Q_{m}$ is the set of
variables used in defining the $m$th rule. A decision rule $r_{m}\in\{0,1\}$
takes the value 1 if all of its $\left|Q_{m}\right|$ conditions are
fulfilled and 0 otherwise. For orderable covariates $s_{k,m}$ will
be an interval or a disjoint union of intervals, while for categorical
covariates $s_{k,m}$ are explicitly enumerated. There is a long tradition
in machine learning to use decision rules as weak learners. Most algorithms
learn decision rules directly from data, such as in \cite{ripper,dembczynski2010ender}.
RuleFit exploits the fact that decision trees can be seen as a set
of decision rules. In a first step a tree ensemble is generated, which
is then decomposed into its defining decision rules. Several efficient
(greedy) algorithmic implementations are available for constructing
the tree ensembles. The generated rules typically correspond to interesting
subspaces with great predictive power. Each node in a decision tree
is defined by a decision rule. Figure~\ref{tree} shows an example tree for
the Boston Housing dataset and Table 1 its corresponding decision
rules. We briefly describe this dataset here since it will be used as a running example
throughout the paper. The Boston housing data consists of $N = 506$ observations which
are city areas in Boston and $p=13$ covariates are recorded.
These variables include ecological measures of nitrogen oxides (NOX),
particulate concentrations (PART) and proximity to the Charles River
(CHAS), the socio-economic variables proportion of black population
(B), property tax rate (TAX), proportion of lower status population
(LSTAT), crime rate (CRIM), pupil teacher ratio (PTRATIO), proportion
of old buildings (AGE), the average number of rooms (RM), area proportion
zoned with large lots (ZN), the weighted distance to the employment
centers (DIS) and an index of accessibility to key infrastructure
(RAD). The dependent variable is the median housing value in the area.
\\
Using Equation \eqref{rule} for example $r_{11}$ can be expressed
as 
\begin{align*}
r_{11}(\bm{x}) & =\prod_{k\in Q_{11}}I(x_{k}\in s_{k,11})=I(6.94\leq RM<7.45)I(LSTAT<9.7).
\end{align*}
This rule is true for areas with relatively large houses with between 6.94 and 7.45 rooms and less than 9.7 \% lower status population.
The $m$th tree consists of $2(u_{m}-1)$ rules, where $u_{m}$ denotes
the number of terminal nodes. Therefore $\sum_{m=1}^{M}2(u_{m}-1)$
rules can be extracted from a tree ensemble of size $M$.\\

\subsection{Collinearity structure of trees}

The generated rules will be combined in a linear model and collinearity
is a concern. For example, the two first child nodes in each tree
are perfectly negative correlated. Furthermore, each parent node is
perfectly collinear with its two child nodes, as it is their union.
One common way to deal with the collinearity problem is to include
the terminal nodes only. This approach also reduces the number of
rules and therefore simplifies computations. We have nevertheless
chosen to consider all possible rules including also non-terminal
ones, but to randomly select one of the two child nodes at each split.
The reason for also including non-terminal nodes is three-fold. First,
even though each parent node in a tree can be reconstructed as a linear
combination of terminal nodes, when using regularization this equivalence
no longer holds. Second, our complexity penalizing prior in Section
3.3 is partly based on the number of splits to measure the complexity
of a rule, and will therefore shrink the several complex child nodes
needed to approximate a simpler parent node. Third, the interpretation
of the model is substantially simplified if the model can select a
simple parent node instead of many complex child nodes. 

\subsection{Generating an informative and diverse rule ensemble}

Any tree method can be used to generate decision rules. Motivated
by the experiments in \cite{friedman2003importance}, Rulefit uses
gradient boosting for rule generation \citep{RuleFit}. Gradient boosting
\citep{friedman2000greedy} fits each tree iteratively on the pseudo
residuals of the current ensemble in an attempt to correct mistakes
made by the previous ensemble. This procedure introduces a lot of
dependence between the members of the ensemble, and many of the produced
rules tend to be informative only when combined to an ensemble. It
might therefore not be possible to remove a lot of the decision rules
without destroying this dependency structure. 

Random Forest on the other hand generates trees independently from
all previous trees \citep{Breiman2001}. Each tree tries to find the
individually best partitioning, given a random subset of observations
and covariates. Random Forest will often generate rules with very
similar splits, and the random selection of covariates forces it to
often generate decision rules based on uninformative predictors. Random
Forest will therefore produce more redundant and uninformative rules
compared to gradient boosting, but the generated rules with strong
predictive power are not as dependent on the rest of the ensemble. 

Since the rules from boosting and Random Forest are very different
in nature, it makes sense to use both types of rules to exploit both
methods' advantages. This naturally leads to a larger number of candidate
rules, but the generalized horseshoe shrinkage proposed in Section
3.2 and 3.3 can very effectively handle redundant rules. Traditional model combination methods
usually use weighting schemes on the output of different  ensemble methods \citep{rokach2010ensemble}. In contrast we combine the extracted rules from the individual trees. To the best of our knowledge this combination of individual weak learners from different ensemble methods is novel and fits nicely in the RuleFit framework with horseshoe regularization, as explained in the Introduction.

The tuning parameters used in the tree generation determine the resulting
decision rules. The most impactful is the tree-depth, controlling
the complexity of the resulting rules. We follow \cite{RuleFit}
with setting the depth of tree $m$ to
\begin{align}
td_{m}=2+\left\lfloor \varphi\right\rfloor 
\end{align}
where $\left\lfloor x\right\rfloor $ is the largest integer less
or equal than $x$ and $\varphi$ is a random variable following the
exponential distribution with mean $L-2$. Setting $L=2$ will produce
only tree stumps consisting of one split. With this indirect specification
the forest is composed of trees of varying depth, which allows the
model to be more adaptive to the data and makes the choice of a suitable
tree depth less important. We use this approach for both boosted and random forest trees.

Another important parameter is the minimum number of observations
in a node $n_{min}$. A too small $n_{min}$ gives very specific rules
and the model is likely to capture spurious relationships. Using $n_{min}=N^{\frac{1}{3}}$
as a default setting has worked well in our experiments, but if prior
information about reasonable sizes of subgroups in the data is available
the parameter can be adjusted accordingly,. Another choice is to determine
$n_{min}$ by cross validation.\\
 In the following all other tuning parameters, e.g the shrinkage parameter
in gradient boosting or the number of splitting covariates in the
Random Forest, are set to their recommended standard choices implemented
in the R-packages \emph{randomForest} and \emph{gbm}.

\section{Ensembles and rule based horseshoe regularization}

This section discusses the \emph{regularization step} of HorseRule
and present a new horseshoe shrinkage prior tailored specifically
for covariates in the form of decision rules.

\subsection{The ensemble}

Once a suitable set of decision rules is generated, they can be combined
in a linear regression model of the form 
\[
y=\alpha_{0}+\sum_{l=1}^{m}\alpha_{l}r_{l}(\mathbf{x})+\epsilon.
\]
As $r_{i}(\mathbf{x})\in\{0,1\}$ they already have the form of dummy
variables and can be directly included in the regression model. A
simple but important extension is to also include linear terms 
\begin{equation}
y=\alpha_{0}+\sum_{j=1}^{p}\beta_{j}x_{j}+\sum_{l=1}^{m}\alpha_{l}r_{l}(\mathbf{x})+\epsilon.\label{rulefit2}
\end{equation}
This extension addresses the difficulty of rule and tree based methods
to approximate linear effects. Splines, polynomials, time effects,
spatial effects or random effects are straightforward extensions of
Equation \eqref{rulefit2}. 

\cite{RuleFit} do not standardize the decision
rules, which puts a higher penalty on decision rules with a smaller
scale. To avoid this behavior, we scale the predictors to have zero
mean and unit variance.

\subsection{Bayesian regularization through the horseshoe prior}

A large set of candidate decision rules is usually necessary to have
a high enough chance of finding good decision rules. The model in
\eqref{rulefit2} will therefore always be high dimensional and often
$p+m>n$. Many of the rules will be uninformative and correlated with
each other. Regularization is therefore a necessity.

RuleFit uses L1-regularized estimates, which corresponds
to an a posterior mode estimator under a double exponential prior
in a Bayesian framework \citep{Tibshirani1996}. As discussed in the
Introduction, the global shrinkage effect of L1-regularization can
be problematic for rule covariates. L1-regularization is well known
to lead to both shrinkage and variable selection. There now exist
implementations of RuleFit that use the elastic net instead of L1-Regularization,
which can lead to improved predictive performance \citep{zou2005regularization},
however elastic net still only uses one global shrinkage parameter.

Another common Bayesian variable selection approach is based on the
spike-and-slab prior \citep{GeorgeMcCulloch1993,Smith1996}
\begin{equation}
\beta_{j}\sim w\cdot N\left(\beta_{j};0,\lambda^{2}\right)+(1-w)\cdot\delta_{0},\label{spike}
\end{equation}
where $\delta_{0}$ is the Dirac point mass function, $N\left(\beta_{j};0,\lambda^{2}\right)$
is the normal density with zero mean and variance $\lambda^{2}$,
and $w$ is the prior inclusion probability of predictor $x_{j}$
. Discrete mixture priors enjoy attractive theoretical properties,
but need to explore a model space of size $2^{(p+m)}$, which can
be problematic when either $p$ or $m$ are large. The horseshoe prior
by \cite{carvalho2009handling,carvalho2010horseshoe} mimics the
behavior of the spike-and-slab but is computationally more attractive.
The regression model with the original horseshoe prior for linear regression is of the form
\begin{align}
y|\textbf{X},\bm{\beta},{\sigma}^{2} & \sim{\mathcal{N}}_{n}(\textbf{X}\bm{\beta},{\sigma}^{2}{\textbf{I}}_{n}),\label{hs}\\
{\beta}_{j}|{\lambda}_{j},{\tau}^{2},{\sigma}^{2} & \sim\mathcal{N}(0,{\lambda}_{j}{\tau}^{2}{\sigma}^{2}),\\
{\sigma}^{2} & \sim{\sigma}^{-2}d{\sigma}^{2},\\
{\lambda}_{j} & \sim{\mathcal{C}}^{+}(0,1),\\
\tau & \sim{\mathcal{C}}^{+}(0,1),
\end{align}
where $\mathcal{C}^{+}(0,1)$ denotes the standard half-Cauchy distribution.
We use horseshoe priors on both linear (the $\beta$'s in Equation \eqref{rulefit2}) and rule terms (the $\alpha$'s in Equation \eqref{rulefit2}). The horseshoe shrinkage for ${\beta}_{j}$ is determined by a local shrinkage parameter ${\lambda}_{j}>0$ and a global shrinkage parameter $\tau>0$. This is important since it allows aggressive shrinking
of noise covariates through small values of $\tau$, while allowing
individual signals to have large coefficients through large ${\lambda}_{j}$.
\citet{carvalho2010horseshoe} show that the horseshoe is better at
recovering signals than the Lasso, and the models obtained from the
horseshoe are shown to be almost indistinguishable from the ones obtained
by a well defined spike-and-slab prior.\\

\subsection{Horseshoe regularization with rule structure}

The original horseshoe assigns the same prior distribution to all
regression coefficients, regardless of the rule's complexity (number
of splits in the tree) and the specificity (number of data points
that fulfill the rule). Similar to the tree structure prior in BART,
we therefore modify the horseshoe prior to express the prior belief
that rules with high length (many conditions) are less likely to reflect
a true mechanism. In addition, we also add the prior information that
very specific rules that are satisfied by only a few data points are
also improbable a priori. The rule support $s(r_{l})\in(0,1)$ is given by $s(r_{j})=N^{-1}\sum_{i=1}^{N}r_{j}(\mathbf{x}_{i})$. Note that a support of 95\% can also be interpreted as 5\%. Therefore we express the specificity of a rule through $\mathrm{min}(1-s(r_{j}),s(r_{j}))$ instead. 
These two sources of prior information are incorporated by extending the prior on $\lambda_{j}$ to 
\[
{\lambda}_{j}\sim{\mathcal{C}}^{+}(0,A_{j}),
\]
with 
\begin{equation}
A_{j}=\frac{\left(2\cdot \mathrm{min}(1-s(r_{j}),s(r_{j}))\right)^{\mu}}{\left(l(r_{j})\right)^{\eta}},\label{eq:rulestructureprior}
\end{equation}
where $l(r_{j})$ denotes the length of rule $j$ defined as its number
of conditions. With increasing number of conditions the prior shrinkage becomes stronger, as well as with increasing specificity. The hyperparameter $\mu$ controls the strength of our belief to prefer general rules that cover a lot of observations and $\eta$ determines
how strongly we prefer simple rules. The response $y$ should be scaled
when using the rule structure prior since the scale of $\beta$ depends
on the scale of $y$.

The rule structure for $A_{j}$ in Equation \eqref{eq:rulestructureprior}
is designed such that $A_{j}=1$ for rules with support $0.5$ and
length $1$, as the ideal. Since $\lim_{\mu\rightarrow0,\eta\rightarrow0}A_{j}=1$
, our rule structure prior approaches the standard horseshoe prior
for small $\mu$ and $\eta$. The rule structure prior gives a gentle
push towards simple and general rules, but its Cauchy tails put considerable
probability mass on non-zero values even for very small $A_{j}$;
the data can therefore overwhelm the prior and keep a complex and
specific rule if needed. 

A model with many complex specific rules may drive out linear terms
from the model, thereby creating an unnecessarily complicated model. We therefore use a standard prior with $A=1$ for linear terms, and set the parameters $\mu$ and $\eta$ to values larger than $0$, which has the effect of giving linear effects a higher chance of being chosen a priori. When $p$ is small it may also be sensible to use no shrinkage at all on the linear effects, and this also allowed in our Gibbs sampling algorithm in Subsection \ref{subsec:PosteriorSampling}.
The hyperparameters $\mu$ and $\eta$ can be chosen guided by theoretical
knowledge about what kind of rules and linear effects are reasonable
for a problem by hand, or determined via cross validation. As a default
choice $(\mu,\eta)=(1,2)$ worked well in our experiments, penalizing
rule complexity heavily and low rule support moderately. 

\subsection{Posterior inference via Gibbs sampling\label{subsec:PosteriorSampling}}

Posterior samples can be obtained via Gibbs sampling. Sampling from
the above hierarchy is expensive, as the full conditionals of $\lambda_{j}$
and $\tau$ do not follow standard distributions and slice sampling
has to be used. \cite{simple} propose an alternative
Horseshoe hierarchy that exploits the following mixture representation
of a half-Cauchy distributed random variable $X\sim{\mathcal{C}}^{+}(0,\varPsi)$
\begin{align}
X^{2}|\psi & \sim\mathcal{IG}\Big(\frac{1}{2},\frac{1}{\psi}\Big),\\
\psi & \sim\mathcal{IG}\Big(\frac{1}{2},\frac{1}{\varPsi^{2}}\Big),
\end{align}
which leads to conjugate conditional posterior distributions. The
sampling scheme in \cite{simple} samples iteratively from the following
set of full conditional posteriors 
\begin{align*}
\bm{\beta}|\cdot & \sim{\mathcal{N}}_{p}({\textbf{A}}^{-1}{\textbf{X}}^{T}\textbf{y},{\sigma}^{2}{\textbf{A}}^{-1})\\
{\sigma}^{2}|\cdot & \sim\mathcal{IG}\Big(\frac{n+p}{2},\frac{(\textbf{y}-\textbf{X}\bm{\beta})^{T}(\textbf{y}-\textbf{X}\bm{\beta})}{2}+\frac{{\bm{\beta}}^{T}{\bm{{\Lambda}}_{\ast}}^{-1}{\bm{\beta}}}{2}\Big)
\end{align*}
\begin{align*}
{\lambda}_{j}^{2}|\cdot & \sim\mathcal{IG}\Big(1,\frac{1}{{\nu}_{j}}+\frac{{{\beta}_{j}}^{2}}{2{\tau}^{2}{\sigma}^{2}}\Big)\\
{\tau}^{2}|\cdot & \sim\mathcal{IG}\Big(\frac{p+1}{2},\frac{1}{{\rho}}+\frac{1}{2{\sigma}^{2}}\sum_{j=1}^{p}\frac{{\beta_{j}}^{2}}{{\lambda_{j}}^{2}}\Big)
\end{align*}
\begin{align*}
\nu_{j}|\cdot & \sim\mathcal{IG}\Big(1,\frac{1}{A^{2}}+\frac{1}{\lambda_{j}^{2}}\Big)\\
\rho|\cdot & \sim\mathcal{IG}\Big(1,1+\frac{1}{\tau^{2}}\Big),
\end{align*}
with $\textbf{A}=({\textbf{X}}^{T}\textbf{X}+{\bm{{\Lambda}}_{\ast}}^{-1})$,
${\bm{\Lambda}}_{\ast}={\tau}^{2}{\bm{\Lambda}}$, ${\bm{\Lambda}}=\text{diag}({{\lambda}_{1}}^{2},\dots,{{\lambda}_{p}}^{2})$.

\subsection{Computational considerations}
The computational complexity of HorseRule can be mainly composed in rule generation and weight learning. The computational cost will thereby always be higher than using boosting or Random Forest alone. This speed disadvantage is partly mitigated by the fact that the HorseRule performs well also without cross-validation.

We have used the R implementations gbm and randomForest here. These algorithms do not scale well for large $N$ and $p$ and become a bottleneck for $N >10000$. This can be easily remedied by migrating the rule generation step to Xtreme Gradient Boosting (XGBoost) \citep{chen2016xgboost} or lightGBM \citep{ke2017lightgbm} that are magnitudes faster for big datasets.

Compared to Bayesian tree learning procedures such as BART or the recently proposed Dirichlet Adaptive Regression Trees (DART) \citep{linero2016bayesian}, no Metropolis-Hastings steps are necessary to learn the tree structure in HorseRule; HorseRule uses only Gibbs sampling on a regularized linear model with rule covariates, which scales linearly with the number of observations \citep{simple}. Sampling 1000 draws from the posterior distribution in the HorseRule model for the Boston housing data used in Section \ref{subsec:Boston-Housing} takes about 90 seconds on a standard computer. The complexity of the Horseshoe sampling depends mostly on the number of linear terms and decision rules, and increases only slowly with $N$. \cite{li2014fully} suggest a computational shortcut where a given $\beta_{j}$ is sampled in a given iteration only if the corresponding scale ($\lambda_{j}\cdot\tau$) is higher than a threshold. The $\lambda_{j}$ needs to be sampled in every iteration to give every covariate the chance of being chosen in the next iteration. We have implemented this approach and seen that it can give tremendous computational gains, but we have not used it when generating the results here since the effects it has on the invariant distribution of the MCMC scheme needs to be explored further. Finally, for very large N (>10000) the linear algebra operations in the Gibbs sampling can become time consuming, and GPU acceleration can be used to speed up sampling \citep{terenin2016gpu}.

\subsection{Sampling the splitting points}

The BART model can be seen as the sum of trees with a Gaussian prior
on the terminal node values
\[
\mu_{j}\sim\mathcal{N}\left(0,\frac{0.5}{\tau\sqrt{k}}\right),
\]
where $k$ denotes the number of trees. BART uses a fixed regularization
parameter $\tau$ and samples the tree structure, while HorseRule
uses a fixed rule structure and adapts to the data through sampling
the shrinkage parameters $\lambda_{j}$ and $\tau$. Using a fixed
tree structure offers dramatic computational advantages, as no Metropolis-Hastings
updating steps are necessary, but the splits are likely to be suboptimal
with respect to the whole ensemble.

As shown in Section \ref{sec:Empirical-results}, both HorseRule
and BART achieve great predictive performance through different means,
and a combination in which both shrinkage and tree structure are sampled
in a fully Bayesian way could be very powerful, but computational
very demanding. An intermediate position is to keep the splitting
variables fixed in HorseRule, but to sample the splitting points.
We have observed that HorseRule often keeps very similar rules with
slightly different splitting points in the ensemble, which is a discrete
approximation to sampling the splitting points. Hence this could also
improve interpretability since a large number of rules with nearby
splitting points can be replaced by a single rule with an estimated
splitting point. It is also possible to replace many similar rules with suitable basis expansions, such as cubic terms or splines.

\section{Results\label{sec:Empirical-results}}
 This section starts out with a predictive comparison of HorseRule against a number of competitors on 16 benchmark datasets. The following subsections explore several different aspects of HorseRule on simulated and real data to evaluate the influence of different components of the model. Subsection \ref{sub:L1vsHorseshoe} contrasts the ability of RuleFit and HorseRule to recover a true linear signal in models with additional redundant rules. The following subsection uses two real datasets to demonstrate the effect of having linear effects in HorseRule, and the advantage of using horseshoe instead of L1 for regularization. Subsection \ref{sub:TwoStepProcedure} addresses that HorseRule uses the training data both to generate the rules and for learning the weights. Subsection \ref{sub:InflRuleGeneratingProcess} explores the role of the rule generating process in HorseRule, and Subsection \ref{sub:NumberOfRules} the sensitivity to the number of rules. Finally in Subsections \ref{sub:Boston} and \ref{sub:cancer} we showcase HorseRule's ability to make interpretable inference from data in different domains.

\subsection{Prediction performance comparison on 16 datasets}\label{sub:16datasets}

We compare the predictive performance of HorseRule with competing
methods on 16 regression datasets. The datasets are a subset of
the datasets used in \cite{0806.3286v2}. From the 23 datasets that were available to us online we excluded datasets that lacked a clear description of which variable to use as response, or which data preprocessing has to be applied to get to the version described in \cite{0806.3286v2}. Since both RuleFit and HorseRule assume Gaussian responses, we also excluded datasets with clearly non-Gaussian response variables, for example count variables with excessive number of zeros. HorseRule can be straightforwardly adapted by using a negative-binomial data augmentation scheme \citep{simple}, but we leave this extension for future work. Table \ref{TableDatasets} displays the characteristics of the datasets.

We compare HorseRule to RuleFit \citep{RuleFit}, Random Forest \citep{Breiman2001}, Bayesian Additive Regression Trees (BART) \citep{0806.3286v2}, Dirichlet Adaptive Regression Trees (DART) \citep{linero2016bayesian}, a recent variant of BART that uses regularization on the input variables, and XGBoost \citep{chen2016xgboost} a highly efficient implementation of gradient boosting.  

We use 10-fold cross validation on each dataset and report the relative RMSE (RRMSE) in each fold;
RRMSE for a fold is the RMSE for a method divided by the RMSE of the
best method on that fold. This allows us to compare performance over
different datasets with differing scales and problem difficulty. We also calculate a worst RRMSE (wRRMSE) on the dataset level, as a measure of robustness. wRRMSE is based on the maximal difference across all datasets between a method's RRMSE and the RRMSE of the best method for that dataset; hence a method with low wRRMSE is not far behind the winner on any dataset. We also calculate the mean RRMSE (mRRMSE) as the relative RMSE on dataset level averaged over all datasets.

To ensure a fair comparison we use another (nested) 5-fold cross validation
in each fold to find good values of the tuning parameters for each
method. For BART and Random Forest the cross-validation settings from
\cite{0806.3286v2} are used. DART is relatively independent of parameter tuning, through the usage of hyperpriors, so we only determine the optimal number of trees. For RuleFit we cross-validate over the number of rules and the depth of the trees, as those are the potentially most impactful parameters. The shrinkage $\tau$ in RuleFit is determined
by the model internally. XGBoost has many parameters that can be optimized, we chose the number of trees, the shrinkage parameter and the treedepth as the most important. For HorseRule we use cross-validation to identify suitable hyperparameters $(\mu, \eta)$ as well as the tree depth. We also run a HorseRule version with the proposed standard settings without cross-validation. Table \ref{TableMethodsSettings} summarizes the settings of all methods.

\begin{table}     
\centering   
\caption{Summary of the 16 regression datasets used in the evaluation.\\$N$, $Q$ and $C$ are the number of observations, quantitative and categorical predictors, respectively.}     
\begin{tabular}{lrrrclrrr}
\toprule 
Name & $N$ & $Q$ & $C$ & & Name & $N$ & $Q$ & $C$\\ 
\midrule 
Abalone & 4177 & 7 & 1 & \hspace{2.7cm} & Diamond & 308 & 1 & 3\\
Ais & 202 & 11 & 1 & & Hatco & 100 & 6 & 4\\
Attend & 838 & 6 & 3 & & Heart & 200 & 13 & 3\\
Baskball & 96 & 4 & 0 & & Fat & 252 & 14 & 0\\
Boston & 506 & 13 & 0 & & Mpg & 392 & 6 & 1\\ 
Budget & 1729 & 10 & 0 & & Ozone & 330 & 8 & 0\\
Cps & 534 & 7 & 3 & & Servo & 167 & 2 & 2\\
Cpu  & 209 & 6 & 1 & & Strike & 625 & 4 & 1\\
\bottomrule 
\end{tabular}\label{TableDatasets}       
\end{table}

\begin{table}
\centering
\caption{Settings for the compared methods.}       
\begin{tabular}{ll}
\toprule    
Method & Parameter settings \\    
\midrule    
HR-default & Ensemble: GBM+RF; $L=5$; $(\mu, \eta) = (1, 2)$. \\    
HR-CV & Ensemble: GBM+RF, $L=(2,5,8)$, $(\mu, \eta) = ((0, 0), (0.5, 0.5), (1, 2))$.\\    
RuleFit &  $k= 500,1000,...,5000$, $L=(2,5,8)$.\\   
Random Forest & Fraction of variables used in each tree = $(0.25,0.5,0.75,1, \sqrt{p}/p)$.\\    
BART & $(\gamma, q)=((3,0.9),(3,0.99),(10,0.75))$; $\tau = 2,3,5$; number of Trees: $50,200$.\\
DART & Number of trees: $50,100$.\\
XGBoost & Number of trees: $50,100,200,350,500$; $\nu = 0(.1,0.05,0.01)$; treedepth: 4,6,8.\\
\bottomrule    
\end{tabular}\label{TableMethodsSettings}     
\end{table}

\begin{figure}[H] \begin{minipage}{\textwidth}   \begin{minipage}[b]{0.49\textwidth} \includegraphics[width=7.5cm, height=7cm]{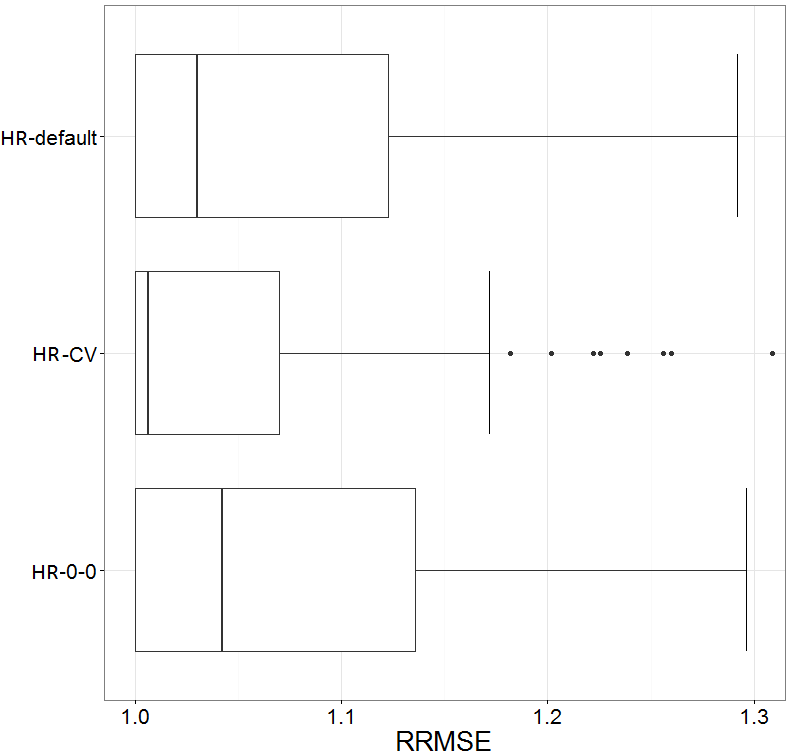} \caption{RRMSE comparison of the different HorseRule versions across all folds.} \label{fig:cv-hs} \end{minipage} \begin{minipage}[b]{0.49\textwidth} \includegraphics[width=7.5cm, height=7cm]{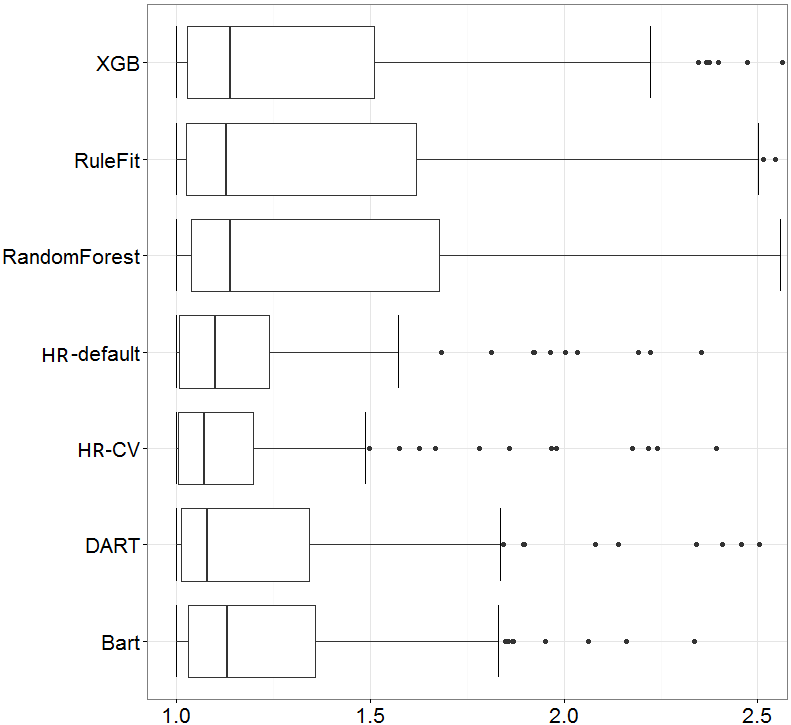}  \caption{RRMSE comparison of HorseRule with competing methods across all folds.} \label{fig:cv-all} \end{minipage} \end{minipage} \end{figure} 

\begin{table}
\centering 
\caption{RRMSE distribution over the 160 crossvalidation folds of the competing methods.}

\begin{tabular}{@{}lcrcrcrcr@{}}

\toprule & 25\%-Quant &  Median & Mean & 75\%-Quant\\ 
\midrule 
\small{}
XGBoost & 1.02 & 1.139 & 1.496 & 1.509\\
RuleFit & 1.026 & 1.129 & 1.426 & 1.618\\ 
RandomForest & 1.039 & 1.137 & 1.508 & 1.677\\
HR-default & 1.007 & 1.101 & \textbf{1.247} & 1.238\\ 
HR-CV & \textbf{1.004} & \textbf{1.072} & 1.262 & \textbf{1.198}\\ 
DART & 1.012 & 1.080 & 1.376 & 1.342\\ 
BART & 1.030 & 1.131 & 1.377 & 1.357\\ 

\bottomrule
\end{tabular} \label{RRMSEdist}
\end{table}

\begin{table}[t] 
\centering 
\caption{Cross-validated prediction performance for the 16 regression datasets. Each entry shows the RMSE and in parentheses the rank on this dataset. The best result is marked in bold.}
\begin{adjustbox}{width=\textwidth}
\begin{tabular}{@{}rcrcrcrcrcr@{}}

\toprule & BART & RForest & RuleFit & HorseRule & HorseRule-CV & DART & XGBoost\\ 
\midrule 
Abalone & 2.150 (7) & 2.119 (3) & 2.139 (5) & 2.115(2) & \textbf{2.114} (1) & 2.129 (4) & 2.147 (6)\\
AIS & 1.144 (4) & 1.247 (7) & 1.207 (6) & 0.713 (2) & \textbf{0.699} (1) & 1.061 (3) & 1.188 (5)\\ 
Attend & 394141 (5) & 411900 (7) & \textbf{345177} (1) & 398485 (6) & 365010 (2) & 370006 (4) & 367231 (3)\\
Baskball & 0.087 (3) & 0.086 (2) & 0.088 (4) & 0.088 (4) & 0.092 (7) & \textbf{0.083} (1) & 0.089 (6)\\ 
Boston & 2.867 (2) & 3.153 (7) & 3.037 (6) & 2.940 (4) & 2.926 (3) & \textbf{2.819} (1) & 2.97 (5)\\ 
Budget & 0.039 (2) & \textbf{0.038} (1) & 0.061 (7) & 0.041 (4) & 0.042 (5) & 0.056 (6) & 0.039 (2)\\ 
CPS & 4.356 (3) & 4.399 (6) & 4.386 (5) & \textbf{4.348} (1) & 4.370 (4) & 4.353 (2) & 4.448 (7)\\ 
Cpu & 41.52 (4) & 54.08 (6) & 54.50 (7) & \textbf{36.03} (1) & 37.47 (3) & 42.87 (5) & 36.75 (2)\\ 
Diamond & 215.0 (3) & 465.9 (7) & 233.7 (4) & 184.5 (2) & \textbf{171.27} (1) & 245.8 (5) & 343.6 (6)\\ 
Hacto & 0.453 (7) & 0.311 (6) & 0.297 (5) & 0.261 (2) & \textbf{0.260} (1) & 0.264 (3) & 0.269 (4)\\ 
Heart & 8.917 (2) & 9.048 (3) & 9.349 (7) & 9.241 (5) & 9.070 (4) & \textbf{8.869} (1) & 9.310 (6)\\ 
Fat & 1.306 (6) & 1.114 (2) & 1.173 (3) & 1.264 (5) & 1.245 (4) & \textbf{1.072} (1) & 1.329 (7)\\ 
MPG & 2.678 (3) & 2.692 (5) & 2.672 (2) & 2.714 (6) & 2.689 (4) & \textbf{2.642} (1) & 2.750 (7)\\ 
Ozone & 4.074 (3) & 4.061 (2) & 4.189 (7) & 4.120 (4) & 4.165 (5) & \textbf{4.054} (1) & 4.174 (6)\\ 
Servo & 0.588 (5) & 0.486 (3) & 0.502 (4) & 0.409 (2) & \textbf{0.403} (1) & 0.671 (6) & 0.719 (7)\\ 
Strikes & 458.4 (7) & 453.7 (5) & 447.7 (3) & 449.2 (4) & 447.2 (2) & \textbf{447.1} (1) & 456.6 (6)\\  
\hline & & & & &\\ Av.Rank & 3.9375 & 4.5625 & 4.8750 & 3.5625 & 3 & \textbf{2.9375} & 5.3125\\
wRRMSE & 1.742 & 2.720 & 1.726 & 1.179 & \textbf{1.160} & 1.666 & 2.006\\
mRRMSE & 1.128 & 1.250 & 1.182 & 1.051 & \textbf{1.035} & 1.141 & 1.201\\
\bottomrule

\end{tabular} \label{tab:cv-regression} \label{TableRMSEAllDatasets}
\end{adjustbox}
\end{table}

We first compare the three different HorseRule versions. Figure \ref{fig:cv-hs} shows the predictive performance of the HorseRule
models over $10\cdot 16 = 160$ dataset and cross-validation splits. While
the $(\mu, \eta) = (1, 2)$ already performs better than the prior
without rule structure ($(\mu, \eta)=(0, 0)$), cross-validation of
$(\mu, \eta)$ helps to improve performance further. 

Figure \ref{fig:cv-all} and Table \ref{RRMSEdist} show that HorseRule has very good performance
across all datasets and folds, and the median RRMSE is 
smaller than its competitors. DART also performs well and is second best in terms of median RRMSE. HorseRule-default is the third best method for the median and best for the mean, which is quite impressive since it does not use cross-validation. 

Table \ref{TableRMSEAllDatasets} summarizes the performance on the dataset level. DART is the best model on $7/16$ datasets and has the best average rank. HorseRule-CV is the best method on $5/16$ datasets and has a slightly worse rank than DART. The last rows of Table \ref{TableRMSEAllDatasets} displays the wRRMSE and mRRMSE over all datasets for each method; it shows
that whenever HorseRule is not the best method, it is only marginally
behind the winner. This is not true for any of the other methods
which all perform substantially worse than the best method on some
datasets. RuleFit performs the best on $1/16$ datasets,
and the median RRMSE is slightly lower than for Random Forest and XGBoost. XGBoost has the hightest median RRMSE and rank in this experiment. This is probably due to the fact, that all methods except Random Forest rely to a certain degree on boosting and improve different aspects of it, making it a hard competition for XGBoost.\\
To summarize, the results show that HorseRule is a highly competitive method with a very stable performance across all datasets. The rule structured prior was found to improve predictive performance, and performs well also without time-consuming cross-validation of its hyperparameters.

\subsection{Regularization of linear terms and rules - RuleFit vs. HorseRule}\label{sub:L1vsHorseshoe}

This subsection uses simulated data to analyse the ability of HorseRule and RuleFit to recover the true signal when
the true relationship is linear and observed with noise. The data
is generated with $X_i \sim \mathcal{N}(0,1), i=1,\dots,100$, $Y = 5X_1 + 3X_2 + X_3 + X_4 + X_5 + \epsilon$ and $\epsilon \sim \mathcal{N}(0,1)$.
The first 5 predictors thus have a positive dependency with $y$ of
varying magnitude while the remaining 95 covariates are noise. Table
\ref{table:TableSimulationStudy} reports the results from 100 simulated
datasets. RMSE measures the discrepancy between the fitted values
and the true mean for unseen test data. RuleFit and HorseRule model use 500 rules in addition
to the linear terms. The best model in RMSE is as expected the Horseshoe
regression without any rules. The OLS estimates without any regularization
struggles to avoid overfitting with all the unnecessary covariates
and does clearly worse than the other methods. HorseRule without the
rule structure prior outperforms RuleFit, but adding a rule structured
prior gives an even better result. The differences between the models
diminishes quickly with the sample size (since the data is rather
clean), the exception being RuleFit which improves at a much lower
rate than the other methods. Table \ref{table:TableSimulationStudy}
also breaks down the results into the ability to recover the true
linear signal, measured by $\Delta \beta_{true} = |(\beta_1 , \beta_2 , \beta_3 , \beta_4 , \beta_5 ) - (5,3,1,1,1)|_1$,
and the ability to remove the noise covariates, measured by $\Delta \beta_{noise} = |(\beta_6 ,\dots, \beta_{100} ) - (0,\dots,0)|_1$.
We see that the HorseRule's horseshoe prior is much better at recovering
the
true linear signal compared to RuleFit with its L1-regularization.
OLS suffers from its inability to shrink away the noise. 

Even though such clear linear effects are rare in actual applications,
the simulation results in Table \ref{table:TableSimulationStudy}
shows convincingly that HorseRule will prioritize and accurately estimate
linear terms when they fit the data well. This is in contrast to RuleFit
which shrinks the linear terms too harshly and compensates the lack
of fit with many rules. HorseRule will only try to add non-linear
effects through decision rules if they are really needed.

\begin{table*}
\centering
\caption{Simulation study. The true effect is linear. }
\label{table:TableSimulationStudy}
\begin{adjustbox}{width=\textwidth}
\begin{tabular}{@{}lrrrcrrrcrrrc@{}}
\toprule 
& \multicolumn{3}{c}{$RMSE$} & \phantom{abc}&
 \multicolumn{3}{c}{$\Delta \bm{\beta}_{true}$} & \phantom{abc}& \multicolumn{3}{c}{$\Delta \bm{\beta}_{noise}$} & \phantom{abc}\\ \cmidrule{2-4} \cmidrule{6-8} \cmidrule{10-12} 
& $n=100$ & $n=500$ & $n=1000$ && $n=100$ & $n=500$ & $n=1000$ && $n=100$ & $n=500$ & $n=1000$\\ 
\midrule 
OLS & 3.23& 1.10& 1.06 && 1.25& 0.19& 0.14&&
2302 & 3.78& 2.54 \\
Horseshoe Regression & 1.14 & 1.01  & 1.01 && 0.40 & 0.18& 0.13 && 
1.72 & 0.70 & 0.49 \\
HorseRule $\alpha=0, \beta = 0$ & 1.54 & 1.02 & 1.01 &&1.99 & 0.39 & 0.29 &&
2.74 & 0.22 & 0.15 \\
HorseRule $\alpha=1, \beta = 2$ & 1.25 & 1.02 & 1.01 && 1.15 & 0.37 & 0.28 &&
3.14 & 0.37 & 0.24 \\
RuleFit $k=2000$ & 1.84 &  1.23 & 1.15 && 3.58 & 1.42 & 1.05 &&
1.18 & 0.91 & 0.99 \\
\bottomrule 
\end{tabular}
\end{adjustbox}

\end{table*}

\subsection{Influence of linear terms in HorseRule, and regularizing by horseshoe instead of L1}\label{sub:InflModelSpec}
\begin{figure}[t!]    \begin{minipage}{\textwidth}  \begin{minipage}[b]{0.49\textwidth} \includegraphics[width=7.5cm, height=7.5cm]{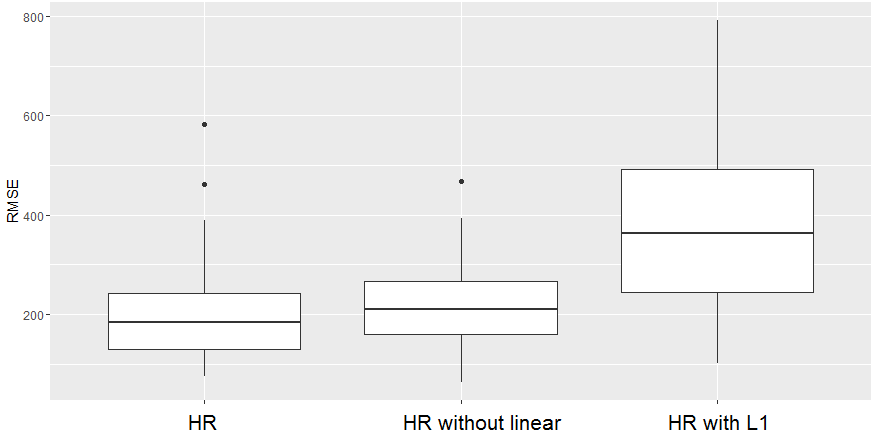}    \end{minipage} \begin{minipage}[b]{0.49\textwidth} \includegraphics[width=7.5cm, height=7.5cm]{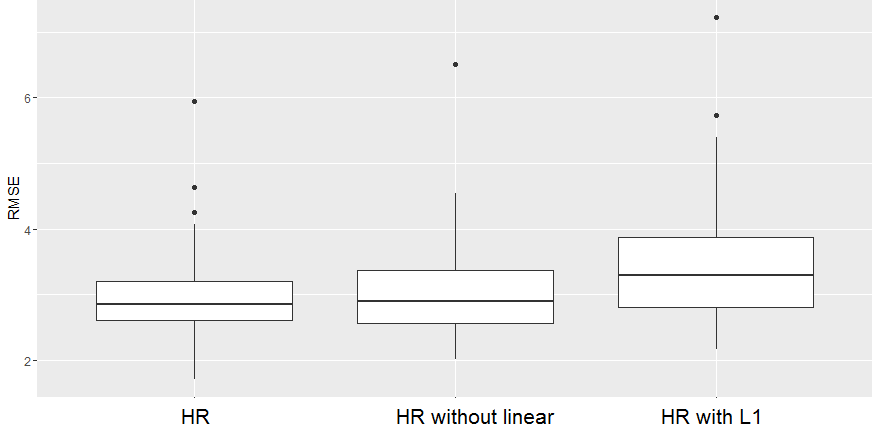}  \end{minipage}
 \caption{RMSE on the Diamonds (left) and the Boston (right) dataset when linear terms are removed and when using L1 regularization instead of horseshoe.} \label{modelspec} \end{minipage} \end{figure}

In this section we analyze to what extent HorseRule's good performance depends on having linear terms in the model, and how crucial the horseshoe regularization is for performance. We demonstrate the effect of these model specification choices on the two datasets Diamonds and Boston. The Diamonds dataset was selected since HorseRule is much better than its competitors on that dataset. The Boston data was chosen since it will be used for a more extensive analysis in Subsection \ref{sub:Boston}. 
Figure~\ref{modelspec} shows the RMSE distribution over the folds used in 10-fold cross-validation. The results are replicated 10 times using different random seeds. The results show that the aggressive shrinkage offered of the horseshoe prior is essential for HorseRule; changing to L1 increases RMSE, especially for the Diamonds data. Note that the L1-version is not entirely identical to RuleFit, as RuleFit uses different preprocessing on rules and only boosting generated rules \citep{RuleFit}. Figure~\ref{modelspec} also shows that adding linear terms gives small decrease of RMSE, but seems less essential for HorseRule's performance.

 \begin{table}
 \centering
 \caption{Median RMSE for different splitting strategies.}
 \begin{tabular}{@{}lcrcrc@{}}\toprule
 & Diamond & Boston & Abalone\\
 All data & 184.6 & 2.851 &  2.115 \\
 50/50 split & 283.7 & 3.555 & 2.136\\
\bottomrule  
\end{tabular} 
\label{split}
\end{table}

\subsection{Influence of the two-step procedure}\label{sub:TwoStepProcedure}
One concern of our two-step procedure is that the same training data is used to find rules and to learn the weights. This double use of the data can distort the posterior distribution and uncertainty estimates. It should be noted however that the rule generation uses only random subsets of data, which mitigates this effect to some extent. It is also important to point out that the predictive results presented in this paper are always on an unseen test set so this is not an issue for the performance evaluations. 

One way to obtain a more coherent Bayesian interpretation is to split the training data in two parts: one part for the rule generation and one part for learning the weights. We can view this as conditionally coherent if the rule learned from the first part of the data is seen as prior experience of the analyst in analyzing the second part of the data. An obvious drawback with such an approach is that less data can be used for learning the model, which will adversely affect predictive performance. Table~\ref{split} displays how predictive performance on the Diamonds ($N=308$) and Boston ($N=506$) data deteriorates from a 50/50 split of the training data. Both these datasets are small and we have also included the moderately large Abalone data ($N=4177$); for this dataset the data splitting has essentially no effect on the performance.  Hence, data-splitting may be an attractive option for moderately large and large data if proper Bayesian uncertainty quantification is of importance.

\subsection{Influence of the rule generating process}\label{sub:InflRuleGeneratingProcess}

In this section we analyze the influence of different rule generating
processes on model performance for the Diamond dataset with ($N=308$
and $p=4$) and the Boston housing data ($N=506$ and $p=13$). \\
In each setting 1000 trees with an average tree depth of $L=5$ are used,
using different ensemble strategies for the rule generation:
\begin{enumerate}
    \item Random Forest generated rules plus linear terms.
    \item Gradient boosting generated rules plus linear terms.
    \item A combination of $30\%$ of the trees from Random Forest and $70$\% from gradient boosting plus linear terms.
\end{enumerate}

\begin{figure}[t!]    \begin{minipage}{\textwidth}  \begin{minipage}[b]{0.49\textwidth} \includegraphics[width=7.5cm, height=7.5cm]{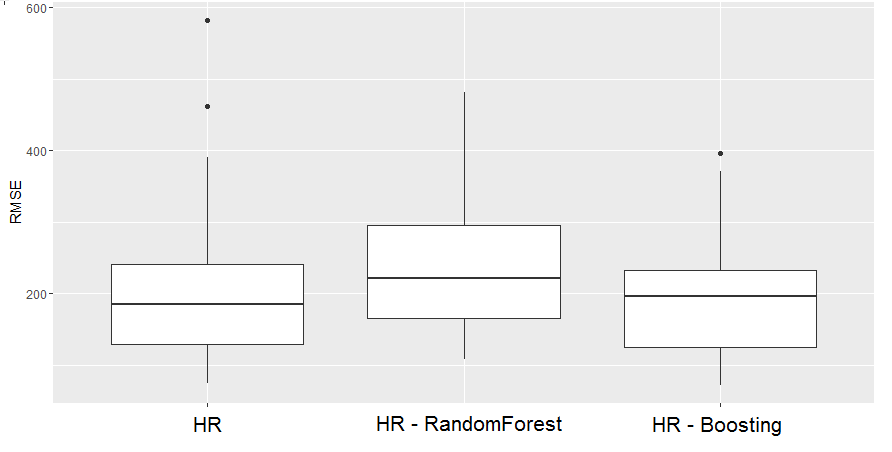}    \end{minipage} \begin{minipage}[b]{0.49\textwidth} \includegraphics[width=7.5cm, height=7.5cm]{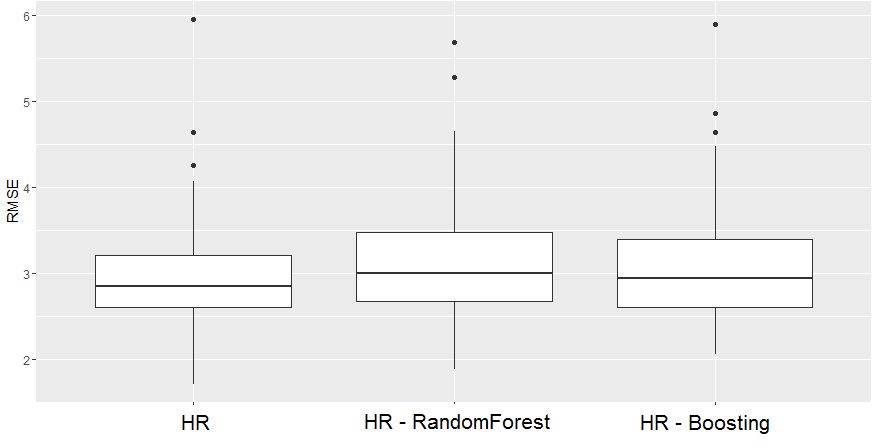}  \end{minipage}
 \caption{RMSE on the Diamonds (left) and the Boston (right) dataset for different rule generating strategies.} \label{Ensemble} \end{minipage} \end{figure} 

The results are shown in Figure~\ref{Ensemble}. As expected the error-correcting
rules found by gradient boosting outperforms randomly generated rules
from Random Forest. However, combining the two types of rules leads to a lower RMSE on both datasets.
In our experiments it rarely hurts the performance to use both type
of rules, and on some datasets it leads to a dramatically better prediction
accuracy. The mixing proportion for the ensemble methods can also
be seen as a tuning parameter to give a further boost in performance. 

\subsection{Influence of the number of rules}\label{sub:NumberOfRules}
\begin{figure}[t!]   \begin{minipage}{\textwidth}   \begin{minipage}[b]{0.5\textwidth} \includegraphics[width=7.5cm, height=7cm]{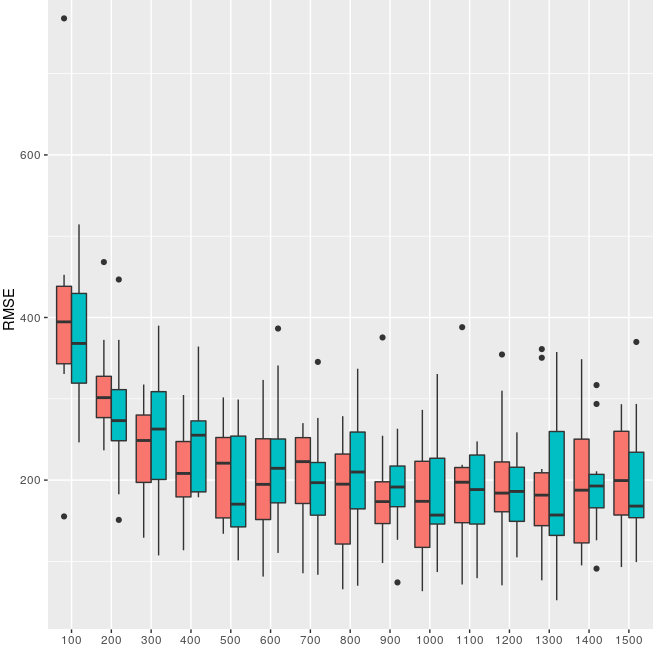} 
 \end{minipage} \begin{minipage}[b]{0.5\textwidth} \includegraphics[width=7.5cm, height=7cm]{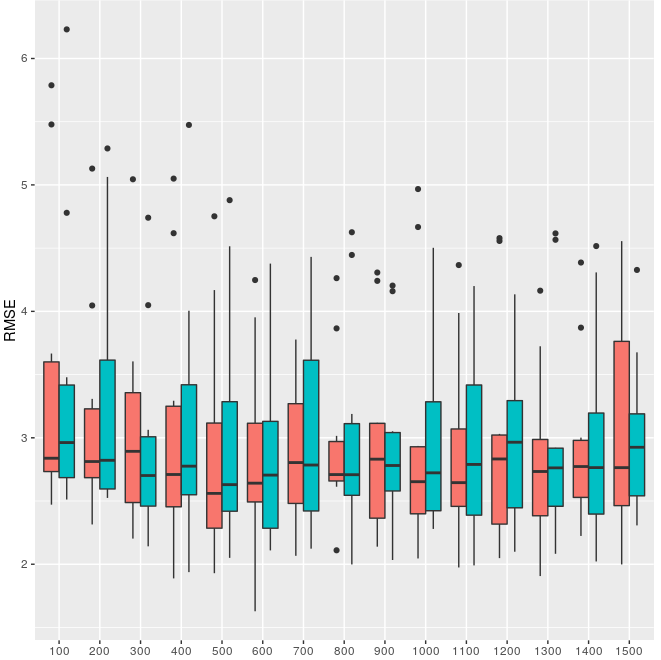} \end{minipage} \caption{RMSE depending on the number of trees on the diamonds (left) and Boston (right) dataset for $(\mu, \eta) = (0,0)$ (red) and $(\mu, \eta) = (1,2)$ (blue).} 
\label{Diamonds} \end{minipage}  \end{figure} 

Another parameter that is potentially crucial is the number of trees
used to generate the decision rules. In gradient boosting limiting
the number of trees (iterations) is the most common way to control
overfitting. Also in BART the number of trees has a major impact on
the quality and performance of the resulting ensemble \citep{0806.3286v2}.
The same is expected for RuleFit, as it uses L1-regularization; with
an increasing number of rules the overall shrinkage $\tau$ increases,
leading to an over-shrinkage of good rules. 

To investigate the sensitivity of HorseRule to the number of trees,
we increase the number of trees successively from 100 to 1500 in the
Boston and Diamonds datasets. This corresponds to ~ $500,550,\dots,5\cdot 1500 =7500$
decision rules before removing duplicates. We also test if the rule
structured prior interacts with the effect of the number of trees
by running the model with $(\mu, \eta) = (0,0)$ and $(\mu, \eta) = (1,2)$.
Figure~\ref{Diamonds} shows the performance of HorseRule as a function
of the number of trees used to extract the rules. Both HorseRule models
are relatively insensitive to the choice of $k$, unless the number
of trees is very small. Importantly, no overfitting effect can be
observed, even when using an extremely large number of 1500 trees
on relatively small datasets ($N=308$ and $N=506$ observations,
respectively). We use 1000 trees as a standard choice, but a small number of trees can be used if computational
complexity is an issue, with little to no expected loss in accuracy.

\begin{minipage}{\linewidth} 
\begin{minipage}[b]{0.6\textwidth}
\begin{figure}[H]
\includegraphics[width=8.5cm,height=8.5cm]{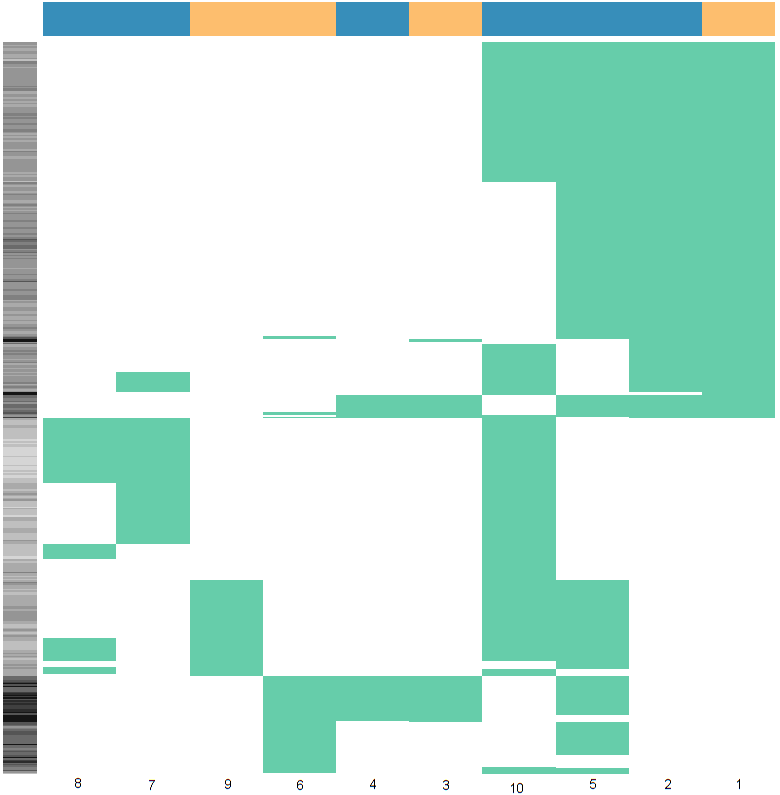} 
\caption{RuleHeat for the Boston Housing data. See the text for details.} \label{RuleHeat}
\end{figure}
\end{minipage}
\begin{minipage}[b]{0.4\textwidth} 

\begin{minipage}[b]{5cm}
\begin{table}[H]

\begin{adjustbox}{width=\linewidth, height=3.65cm} \begin{tabular}{@{}llrcrcrcr@{}}
\toprule 
Rule & & $5\%\,I$ & $\bar{I}$ & $95\%\,I$ & $\bar{\beta}$\\ 
\midrule 
1 & $RM \leq 6.97$ & 0.96 & 0.99 & 1.00 & \hspace{0.1cm}24.1 \\
& $ LSTAT \leq 14.4$ & & &\\
2 & $RM \leq 6.97 $  & 0.77 & 0.89 & 1.00 & -21.9 \\ 
 & $ DIS > 1.22$  & & &\\
 & $LSTAT \leq 14.4$ & & &\\
3 & $LSTAT \leq 4.66$ & 0.00 & 0.27 & 0.51 & \hspace{0.1cm}12.35 \\
4&$TAX \leq 416.5 $  & 0.00 & 0.21 & 0.43 & -10.46\\ 
& $LSTAT \leq 4.65$ & & &\\
5&$NOX \leq 0.59$ & 0.00 & 0.12 & 0.21 & -2.94 \\ 
6&$NOX \leq 0.67$  & 0.00 & 0.10 & 0.33 & \hspace{0.1cm}3.87 \\
& $RM > 6.94$ & & &\\
7&$NOX > 0.67$ & 0.00 & 0.11 & 0.37 & -3.24\\ 
8&$LSTAT > 19.85 $ & 0.00 & 0.15 & 0.53 & -3.18\\ 
9&$linear: AGE $ & 0.00 & 0.09 & 0.15 & -0.03\\ 
10&$linear: RAD$ & 0.00 & 0.07 & 0.19 & \hspace{0.1cm}0.10\\ 
\bottomrule 
\end{tabular}
\end{adjustbox}
\end{table}
\end{minipage}
\begin{minipage}[b]{6cm}
\begin{table}[H]
\caption{The ten most important rules in the Boston housing data.} \label{TenRulesBoston}
\end{table}
\end{minipage}
\end{minipage} \end{minipage}

\vspace{1.5cm}

\subsection{Boston Housing\label{subsec:Boston-Housing}}\label{sub:Boston}

In this section we apply HorseRule to the well known Boston Housing
dataset to showcase its usefulness in getting insights from the data. For a detailed description of the dataset see Section~\eqref{sec:rule-generation}.\\
The HorseRule with default parameter settings is used to fit the model.
Table \ref{TenRulesBoston} shows the 10 most important effects. Following
\cite{RuleFit}, the importance of a linear term is defined as
\[
I(x_{j})=\left|\text{\ensuremath{\beta}}_{j}\right|sd(x_{j})
\]
where $sd(\cdot$) is the standard deviation, and similarly for a
predictor from a decision rule
\[
I(r_{l})=\left|\text{\ensuremath{\alpha}}_{l}\right|sd(r_{l}).
\]
We use the notation $I_{j}$ when it is not important to distinguish
between a linear term and a decision rule. For better interpretability
we normalize the importance to be in $[ 0, 1]$, so that the most
important predictor has an importance of 1. Table~\ref{TenRulesBoston} reports the posterior
distribution of the normalized importance (obtained from the MCMC
draws) of the 10 most important rules or linear terms. The most important
single variable is LSTAT, which appears in many of the rules, and
as a single variable in the third most important rule. Note also that
LSTAT does not appear as a linear predictor among the most important
predictors. 

\begin{figure}[t!]  \includegraphics[width=10cm, height=8cm]{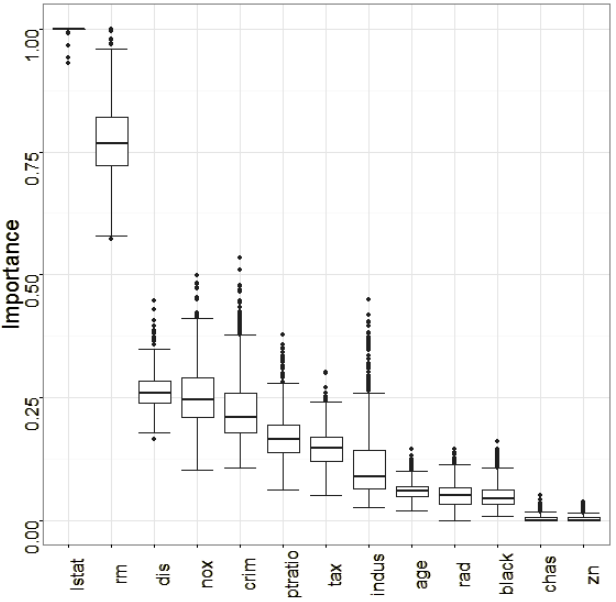}  \caption{Posterior distribution of the Input Variable Importance for the 13 Covariates.}\label{bostonimportance} \end{figure} 

To interpret the more complex decision rules in Table \ref{TenRulesBoston}
it is important to understand that decision rules in an ensemble have
to be interpreted with respect to other decision rules, and in relation
to the data points covered by a rule. A useful way to explore the
effects of the most important rules is what we call a \emph{RuleHeat}
plot, see Figure \ref{RuleHeat} for an example for the Boston housing
data. The horizontal axis lists the most important decision rules
and the vertical axis the $N$ observations. A square is green if
$r_{l}(\mathbf{x})=1$. The grayscale on the bar to the left indicates
the outcome (darker for higher price) and the colorbar in the top
of the figure indicates the sign of the covariate's coefficient in
the model (sand for positive). RuleHeat makes it relatively easy to
to find groups of similar observations, based on the rules found in
HorseRule, and to assess the role a rule plays in the ensemble. For
example, Figure \ref{RuleHeat} shows that the two most important
rules differ only in a few observations. The two rules have very large
coefficients with opposite signs. Rule 1 in isolation implies that
prices are substantially higher when the proportion of lower status
population is low (LSTAT$\leq14.4$) for all but the very largest
houses (RM$\leq6.97$). However, adding Rule 2 essentially wipes out
the effect of Rule 1 ($24.1-21.9 = 2.2$) except for the six houses
very close to the employment centers (DIS<1.22) where the effect on
the price remains high. 

Similarly to the Variable importance in Random Forest and RuleFit,
we can calculate a variable input importance for the HorseRule model.
The importance of the $j$th predictor given the data is defined as
\citep{RuleFit} 
\[
J(x_{j})=I(x_{j})+\sum_{l:j\in Q_{l}}I(r_{l})/\left|Q_{l}\right|
\]
where the sum runs over all rules where $x_{j}$ is one of the predictors
used to define the rule. Note how the importance of the rules are discounted
by the number of variables involved in the rule, $\left|Q_{l}\right|$.
Figure \ref{bostonimportance} shows the
posterior distribution of $J(x_{j})$ for the 13 covariates. LSTAT
is the most important covariate with median posterior probability
of 1 and very narrow posterior spread, followed by RM which has a
median posterior importance of around 0.75. The importance of some
variables, like NOX and INDUS, has substantial posterior uncertainty
whereas for other covariates, such as AGE, the model is quite certain
that the importance is low (but nonzero).

\begin{table}
\caption{The ten most important rules in Boston data after DSS.} \label{DSSBoston}
\begin{adjustbox}{width=5cm, height=3.5cm} 

\begin{tabular}{@{}lllcrcrcr@{}}
\toprule 
Rule & & $\bar{I}$ & $\bar{\beta}$\\ 
\midrule 
1 & $RM \leq 7.13$ & 1.00 & -3.47\\
2 & $RM \leq 6.98 $  & 0.97 & -2.36\\ 
 & $PTRATIO \leq 18.7$ &\\
 & $LSTAT > 5.95$ & & &\\
3 & $LSTAT > 18.75 $ & 0.81 & \hspace{0.2cm}1.80 \\
4&$linear: RAD $ & 0.80& \hspace{0.1cm}0.10\\ 
5&$RM \leq 7.437  $ & 0.79 &-2.03\\ 
 & $LSTAT \leq 7.81$ &\\
6&$NOX \leq 0.62 $ & 0.70 & -1.64\\
 & $RM  \leq 7.31$ &\\
7&$RM \leq 7.1$ & 0.68 & -2.47\\ 
 & $RAD \leq 4.5 $ &\\
 & $LSTAT \leq 7.81$ &\\
8&$NOX > 0.59 $ & 0.63 & -1.47\\ 
9&$linear: LSTAT $ & 0.58 & -0.09\\ 
10&$linear: AGE $ & 0.58 & -0.02\\ 
\bottomrule 
\end{tabular}
\end{adjustbox}
\end{table}

The overlapping rules, as well as similar rules left in the ensemble
in order to capture model uncertainty about the splitting points make
interpretation somewhat difficult. One way to simplify the output
from HorseRule is to use the \textit{decoupling shrinkage and summary
}(DSS) approach by \cite{hahn2015decoupling}.
The idea is to reconstruct the full posterior estimator $\hat{\beta}$
with a 1-norm penalized representation, that sets many of the coefficients
to exactly zero and also merges together highly correlated coefficients.
We do not report systematic tests here, but in our experiments using
DSS with a suitable shrinkage parameter did not hurt the predictive
performance, while allowing to set a vast amount of coefficients to
zero. Using HorseRule followed by DSS on the Boston housing data leaves
106 non-zero coefficients in the ensemble. The 10 most important rules
can be seen in Table \ref{DSSBoston}. We can see that the new coefficients
are now less overlapping. The relatively small number of rules simplify
interpretation. Posterior summary for regression with shrinkage priors
is an active field of research (see e.g. \cite{piironen2016comparison} and \cite{puelz2016variable} for interesting approaches)
and future developments might help to simplify the rule ensemble further.

\subsection{Logistic Regression on gene expression data}\label{sub:cancer}

Here we analyze how HorseRule can find interesting
pattern in classification problems, specifically in using gene expression data for finding genes that can signal the presence
or absence of cancer. Such information is extremely important since it can be used to construct explanations about the underlying biological mechanism that lead to mutation, usually in the form of gene pathways. Supervised gene expression classification can also help to design diagnostic tools and patient predictions, that help to identify the cancer type in early stages of the disease and to decide on suitable therapy \citep{van2002gene}. 

\begin{table}
\caption{Accuracy in training and test set for the prostate cancer data} \label{AccuracyGene}

\begin{tabular}{@{}lrcrcrcrcrcr@{}}
\toprule & BART & Random Forest & RuleFit & HorseRule\\ 
\midrule 
CV-Accuracy & 0.900 & 0.911 & 0.831 & 0.922\\
CV-AUC & 0.923 & 0.949 & 0.953 & 0.976\\ 
Test-Accuracy & 0.824 & 0.971 & 0.941 & 0.971\\
Test-AUC & 1 & 0.991 & 0.995 & 1\\ 
\bottomrule

\end{tabular} 

\end{table}

Extending HorseRule to classification problems can be easily done using a latent variable formulation of, for example, the logistic regression. We chose
to use the P{\'o}lya--Gamma latent variable scheme by \cite{polson2013bayesian}.
Methodological difficulties arise from the usually small number of
available samples, as well as high number of candidate genes, leading
to an extreme $p>>n$ situation. We showcase the ability of HorseRule
to make inference in this difficult domain on the Prostate Cancer
dataset, which consists of 52 cancerous and 50 healthy samples $(n=102)$. In
the original data $p=12600$ genetic expressions are available, which
can be reduced to 5966 genes after applying the pre-processing described
in \citet{singh2002gene}. 
Since spurious relationships can easily
occur when using higher order interactions in the $p >> n$ situation,
we use the hyperparameters $\mu=2$ and $\eta=4$ to express our prior
belief that higher order interactions are very unlikely to reflect
any true mechanism. 

Table~\ref{AccuracyGene} shows that HorseRule has higher accuracy
and significantly higher AUC than the competing methods. We also test
the methods on an unseen test dataset containing 34 samples not used
in the previous step. All methods have lower error here, implying
that the test data consists of more predictable cases. The difference
is smaller, but HorseRule performs slightly better here as well.

The 10 most important rules for HorseRule are founds in Table \ref{ImportantRulesGene}. It contains 8 rules with one condition and only 2 with two conditions,
implying that there is not enough evidence in the data for complicated
rules to overrule our prior specification. All of the most important
rules still contain 0 in their $5\%$ posterior importance distribution,
implying that they are eliminated by the model in at least $5\%$
of the samples; the small sample size leads to non-conclusive results. 

\begin{minipage}{\linewidth} 
\begin{minipage}[b]{0.6\textwidth}
\begin{figure}[H]
\includegraphics[width=8.5cm,height=8.5cm]{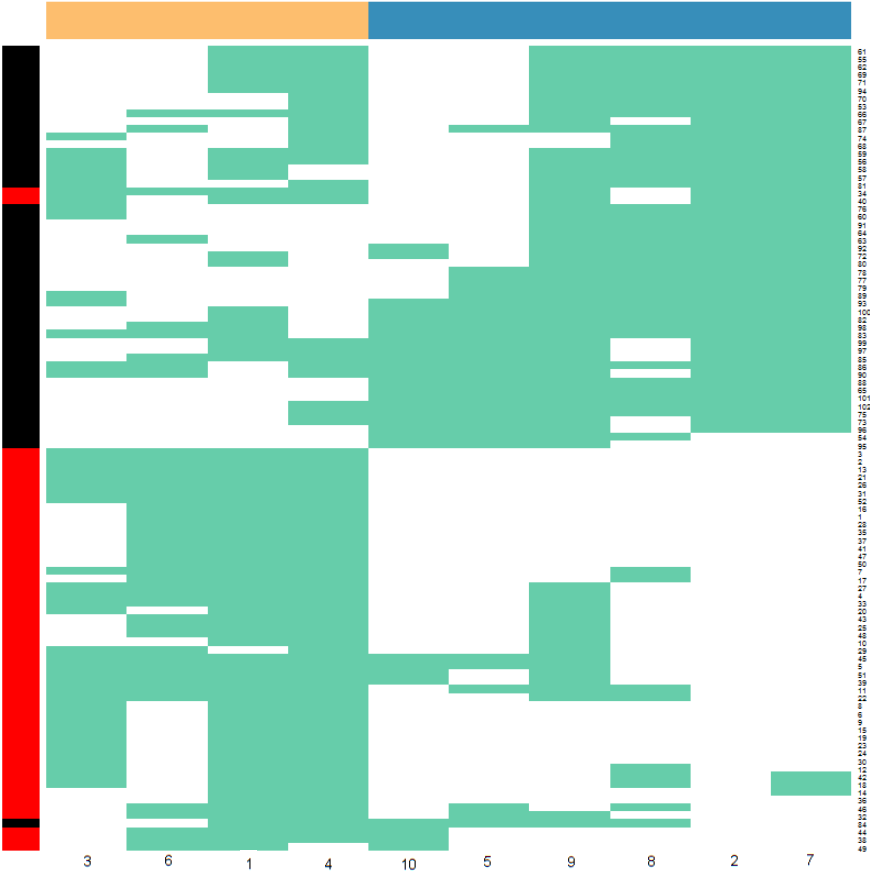} 
\caption{RuleHeat for the prostate cancer data. Cancer tissues are colored in red, healthy in black.} \label{RuleHeatGene}
\end{figure}
\end{minipage}
\begin{minipage}[b]{0.4\textwidth} 

\begin{minipage}[b]{5cm}
\begin{table}[H]

\begin{adjustbox}{width=\linewidth, height=3.1cm} \begin{tabular}{@{}lllcrcrcr@{}}
\toprule 
Rule & & $5\%\,I$ & $\bar{I}$ & $95\%\,I$ & $\bar{\beta}$\\ 
\midrule 
1 & $556\_s\_at \leq 1.55$ & 0 & 0.33 & 1 & 3.10 \\
2 & $34647\_at \leq -1.18 $  & 0 & 0.15 & 1 & -1.78 \\ 
 & $ 37639\_at \leq 1$ & & &\\
3 & $37478 > -0.32$ & 0 & 0.18 & 0.91 & 1.42 \\
4&$38087\_s\_at \leq 0.83$ & 0 & 0.23 & 1 & 1.81\\ 
5&$34678\_at > 0.38$ & 0 & 0.19 & 0.88 & -1.58 \\ 
6&$1243\_at \leq 0.35$ & 0 & 0.15 & 0.66 & 1.19 \\ 
7&$37639\_at \leq 1$ & 0 & 0.13 & 0.80 & -1.10\\ 
8&$33121\_g\_at \leq 0.672 $ & 0 & 0.10 & 0.82 & -1.09 \\ 
& $  960\_g\_at > 0.378$ & & & \\
9&$41706\_at \leq 1.33 $ & 0 & 0.15 & 0.79 & -1.13 \\ 
10&$39061\_at > 0.31$ & 0 & 0.1 & 0.52 & -1.03 \\ 
\bottomrule 
\end{tabular}
\end{adjustbox}
\end{table}
\end{minipage}
\begin{minipage}[b]{6cm}
\begin{table}[H]
\caption{Ten most important rules in the cancer data.} \label{ImportantRulesGene}
\end{table}
\end{minipage}
\end{minipage} \end{minipage}

\begin{figure}[t]   \begin{minipage}{\textwidth}  \begin{minipage}[b]{0.49\textwidth} \includegraphics[width=7.4cm, height=7.5cm]{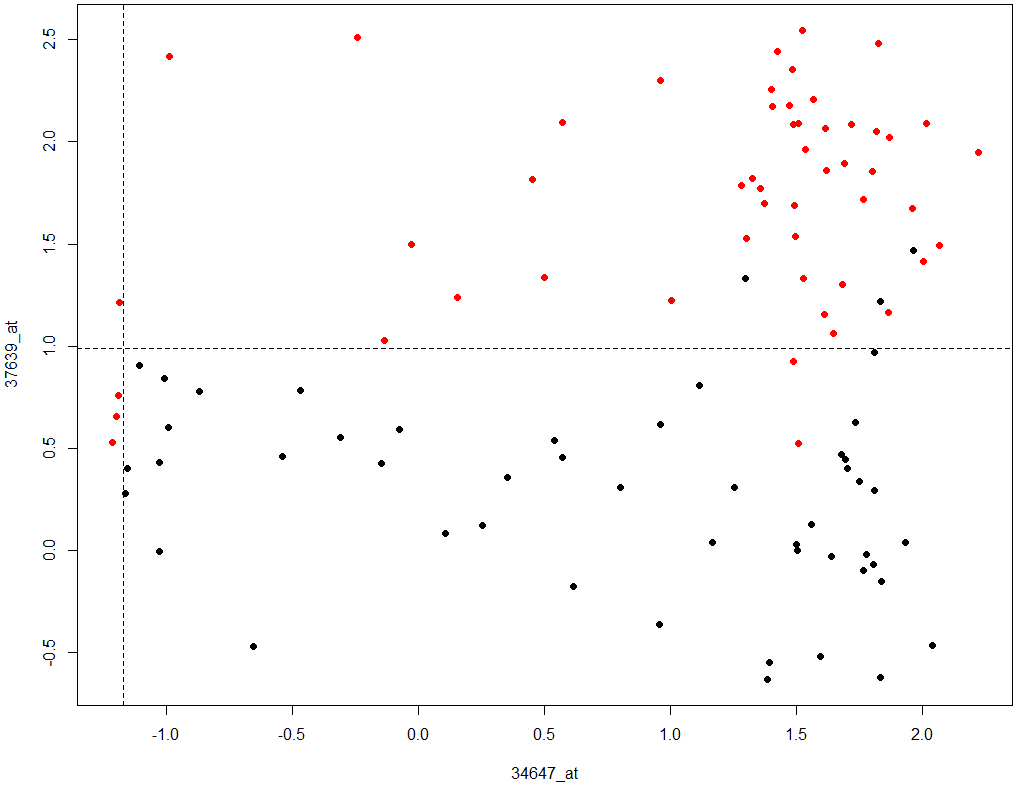} \caption{Scatterplot for Genes $37639\_at$ and $34647\_at$. Healthy samples in black and cancerous samples in red. Rule 2 is defined by the bottom right quadrant.} \label{ScatterplotGenes1} \end{minipage} \begin{minipage}[b]{0.49\textwidth} \includegraphics[width=7.4cm, height=7.5cm]{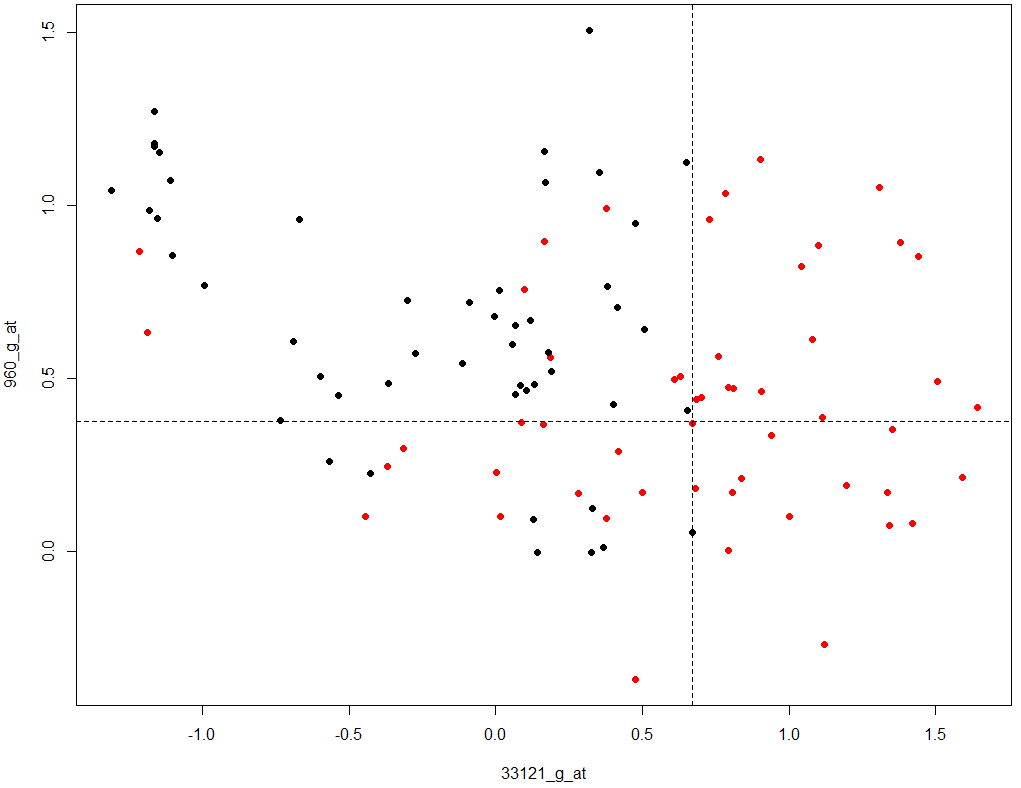} \caption{Scatterplot for Genes $33121\_g\_at$ and $960\_g\_at$. Healthy samples in black and cancerous samples in red. Rule 8 is defined by the top left quadrant.} \label{ScatterplotGenes2} \end{minipage}  \end{minipage} \end{figure} 

\begin{figure}[t]  \includegraphics[width=12cm, height=8cm]{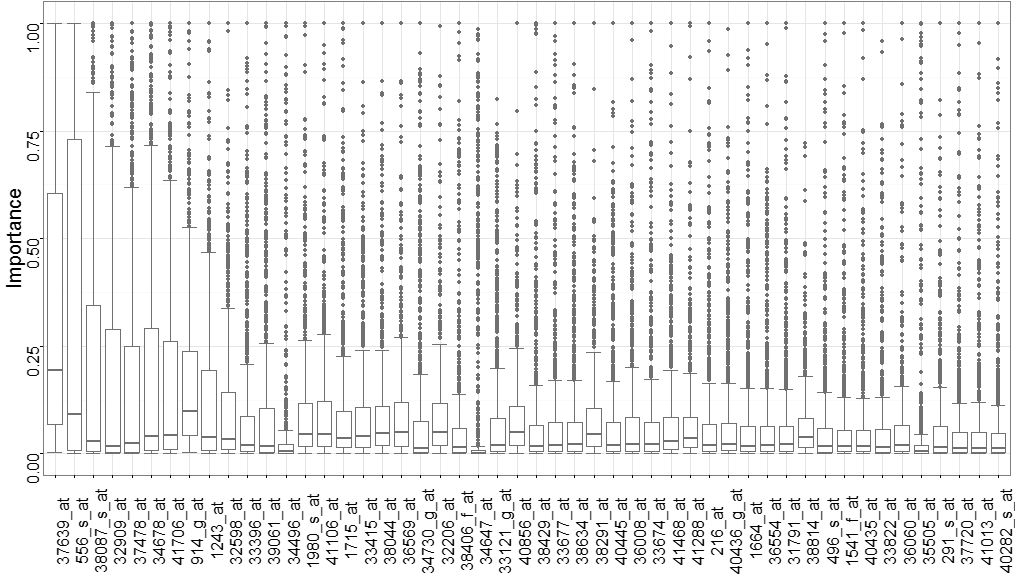} \label{fig:gene-importance} \caption{Posterior distribution of the Input Variable Importance of the 50 most influential Covariates.} \label{VariableImportanceGenes} \end{figure} 
Figure \ref{VariableImportanceGenes} shows the input variable importance of the 50 most important genes. In this domain the advantage of having
estimates of uncertainty can be very beneficial, as biological follow
up studies are costly and the probability of spurious relationships
is high. In this data the genes $37639\_{at}$ and $556\_s\_{at}$
contain an importance of 1 in their 75 \% posterior probability bands. The gene $37639\_{at}$ was found in previous studies to
be the single gene most associated with prostate cancer \citep{yap2004classification}. However, gene $556\_s\_{at}$, which makes up the most important Rule 1, was only found to be the 9th important in previous studies on the same data using correlation based measures \citep{yap2004classification}. So, while this gene is individually not very discriminative (77\% accuracy), it becomes important in conjunction with other rules. This is also borne out in the RuleHeat plot in Figure \ref{RuleHeatGene}. The outcome is binary, and the vertical bar to the
left is red for cancer and black for healthy. RuleHeat shows that Rule
1 covers all except one cancer tissue together with a number of normal tissues, and would therefore probably not
be found to be significant using traditional tests in logistic regression. Its importance arises
from the combination with the other rules, especially Rule 2, Rule
7 and Rule 8, that are able to correct the false positive predictions
using Rule 1 alone.

To illustrate HorseRule's potential for generating important insights from interaction rules, we present the
subspaces of the two most important interaction rules in  Figure~\ref{ScatterplotGenes1} and Figure~\ref{ScatterplotGenes2}. Again healthy
tissues are colored black and cancerous red. The first interaction
looks somewhat unnatural. The gene $37639\_at$ is individually seen to be a strong classifier where higher values indicate cancer. This rule is also individually
represented as Rule 7. The second split on $34647\_at <-1.18$ corrects
3 misclassified tissues by the first split alone. This rule
probably only works well in the ensemble but may not reflect a true
mechanism. The second interaction effect is more interesting. It seems that healthy
tissues have lower values in the expression of $33121\_g\_at$ and higher values in
the expression of $960\_g\_at$. This rule might reflect a true interaction mechanism
and could be worth analysing further.

Overall, this shows that HorseRules non-linear approach with interacting rules complements the results from classical linear approaches with new information. Decision rules are especially interesting for the construction of gene-pathways \citep{glaab2010learning}, diagnostic tools and identification of targets for interventions \citep{slonim2002patterns}.

\section{Conclusions}

We propose HorseRule, a new model for flexible non-linear regression
and classification. The model is based on RuleFit and uses decision
rules from a tree ensemble as predictors in a regularized linear fit.
We replace the L1-regularization in RuleFit with a horseshoe prior
with a hierarchical structure especially tailored for a situation
with decision rules as predictors. Our prior shrinks complex (many
splits) and specific (small number of observations satisfy the rule)
rules more heavily a priori, and is shown to be efficient in removing
noise without tampering with the signal. The efficient shrinkage properties
of the new prior also makes it possible to complement the rules from
boosting used in RuleFit with an additional set of rules from random
forest. The rules from Random Forest are not as tightly coupled as
the ones from boosting, and are shown to improve prediction performance
compared to using only rules from boosting. 

HorseRule is shown to outperform state-of-the-art competitors
like RuleFit, BART and Random Forest in an extensive evaluation of
predictive performance on 16 widely used datasets. Importantly, HorseRule
performs consistently well on all datasets, whereas the other methods
perform quite poorly on some of the datasets. We explored different aspect of HorseRule to determine the underlying factors behind its success. We found that the combination of mixing rule from different tree algorithms and the aggressive but signal-preserving horseshoe shrinkage are essential, but that the addition of linear terms seems less important. Our experiments also show that the predictive performance of HorseRule is not sensitive to its prior hyperparameters. We also demonstrate the
interpretation of HorseRule in both a regression and a classification
problem. HorseRule's use of decision rules as predictors and its ability
to keep only the important predictors makes it easy to interpret its
results, and to explore the importance of individual rules and predictor
variables.

\section*{Appendix A - The HorseRule R package}

The following code illustrates the basic features of our HorseRule
package in R with standard settings. The package is available on CRAN at \url{https://CRAN.R-project.org/package=horserule}. \texttt{\medskip{}
}

\texttt{\textbf{\footnotesize{}library(horserule)}}{\footnotesize \par}

\texttt{\textbf{\footnotesize{}data(Boston, package = "MASS" )
}}\texttt{\footnotesize{}\medskip{}
}{\footnotesize \par}

\texttt{\textbf{\footnotesize{}N = nrow(Boston)}}{\footnotesize \par}

\texttt{\textbf{\footnotesize{}train = sample(1:N, 500)}}{\footnotesize \par}

\texttt{\textbf{\footnotesize{}Xtrain = Boston{[}train,-14{]}}}{\footnotesize \par}

\texttt{\textbf{\footnotesize{}ytrain = Boston{[}train, 14{]}}}{\footnotesize \par}

\texttt{\textbf{\footnotesize{}Xtest = Boston{[}-train, -14{]}}}{\footnotesize \par}

\texttt{\textbf{\footnotesize{}ytest = Boston{[}-train, 14{]}}}\texttt{\footnotesize{}\medskip{}
}{\footnotesize \par}
\texttt{\footnotesize{}\# Selecting predictors to be included as linear terms}{\footnotesize \par}

\texttt{\textbf{\footnotesize{}lin = 1:13}}{\footnotesize \par}

\texttt{\footnotesize{}\# Main function call (variable scaling performed
internally)}{\footnotesize \par}

\texttt{\textbf{\footnotesize{}hrres = HorseRuleFit(X=Xtrain, y=ytrain,}}{\footnotesize \par}

\texttt{\footnotesize{}$\:$\qquad{}\qquad{}\# MCMC settings }{\footnotesize \par}

\texttt{\footnotesize{}$\:$\qquad{}\qquad{}}\texttt{\textbf{\footnotesize{}thin=1,
niter=1000, burnin=100,}}{\footnotesize \par}

\texttt{\footnotesize{}$\:$\qquad{}\qquad{}\# Parameters for the
rule generation process}{\footnotesize \par}

\texttt{\footnotesize{}$\:$\qquad{}\qquad{}}\texttt{\textbf{\footnotesize{}L=5,
S=6, ensemble = \textquotedbl{}both\textquotedbl{}, mix=0.3, ntree=1000,}}{\footnotesize \par}

\texttt{\footnotesize{}$\:$\qquad{}\qquad{}\# Model parameters. Data
is scaled so no intercept needed.}{\footnotesize \par}

\texttt{\footnotesize{}$\:$\qquad{}\qquad{}}\texttt{\textbf{\footnotesize{}intercept=F,
linterms=lin, ytransform = \textquotedbl{}log\textquotedbl{},}}{\footnotesize \par}

\texttt{\footnotesize{}$\:$\qquad{}\qquad{}\# Hyperparameters for
the rule structured prior}{\footnotesize \par}

\texttt{\footnotesize{}$\:$\qquad{}\qquad{}}\texttt{\textbf{\footnotesize{}alpha=1,
beta=2, linp = 1, restricted = 0)}}{\footnotesize \par}

\texttt{\footnotesize{}\smallskip{}
}{\footnotesize \par}

\texttt{\footnotesize{}\# Check model performance by predicting holdout
cases}{\footnotesize \par}

\texttt{\textbf{\footnotesize{}pred = predict(hrres, Xtest)}}{\footnotesize \par}

\texttt{\textbf{\footnotesize{}sqrt(mean((pred-ytest)\textasciicircum{}2))}}{\footnotesize \par}

\texttt{\footnotesize{}\smallskip{}
}{\footnotesize \par}

\texttt{\footnotesize{}\# Find most important rules}{\footnotesize \par}

\texttt{\textbf{\footnotesize{}importance\_hs(hrres)}}{\footnotesize \par}

\texttt{\footnotesize{}\medskip{}
}{\footnotesize \par}

\texttt{\footnotesize{}\# Compute variable importance}{\footnotesize \par}

\texttt{\textbf{\footnotesize{}Variable\_importance(hrres)}}{\footnotesize \par}

\texttt{\footnotesize{}\medskip{}
}{\footnotesize \par}

\section*{Acknowledgements}
We are grateful to the two reviewers and the associate editor for constructive comments that helped to improve both the presentation and the contents of the paper.

\bibliographystyle{plainnat}
\bibliography{bib}
\end{document}